\documentclass[aps,prl,showkeys]{revtex4}

\usepackage{graphicx}
\usepackage{amssymb}
\usepackage{amsfonts,amsmath}

\begin{document}

\title{Optical control of scattering, absorption and lineshape in
  nanoparticles}
\author{Benjamin Hourahine, Francesco Papoff}  
\affiliation{Department of Physics,
 University of Strathclyde, 107 Rottenrow, Glasgow G4 0NG, UK. (email: f.papoff@strath.ac.uk)}


\begin{abstract}

We find exact conditions for the enhancement or suppression of
internal and/or scattered fields and the determination of their
spatial distribution or angular momentum through the combination of
simple fields.  The incident fields can be generated by a single
monochromatic or broad band light source, or by several sources, which
may also be impurities embedded in the nanoparticle.  We can design
the lineshape of a particle introducing very narrow features in its
spectral response.
\end{abstract}

\maketitle

Control of near and far field optical emission and the optimization of
coupling between incident light and nanostructures are fundamental
issues that underpin the ability to enhance sub-wavelength linear and
nonlinear light-matter interaction processes in
nanophotonics. Nonlinear~\cite{abb11a} and linear control based on
pulse shaping~\cite{kubo05a,durach07a}, combination of adaptive
feedbacks and learning algorithms~\cite{aeschlimann07a}, as well as
optimization of coupling through coherent absorption~\cite{noh12a} and
time reversal~\cite{pierrat13a} have all been recently investigated.
Interference between fields was some time ago proposed in quantum
optics as a way to suppress losses in lossy beam
splitter~\cite{jeffers00a} and has been recently applied to show
control of light with light in linear plasmonic
metamaterials~\cite{zhang12a}.

\begin{figure}
  \begin{center}a)
    \includegraphics[clip=true,width=.25\textwidth]{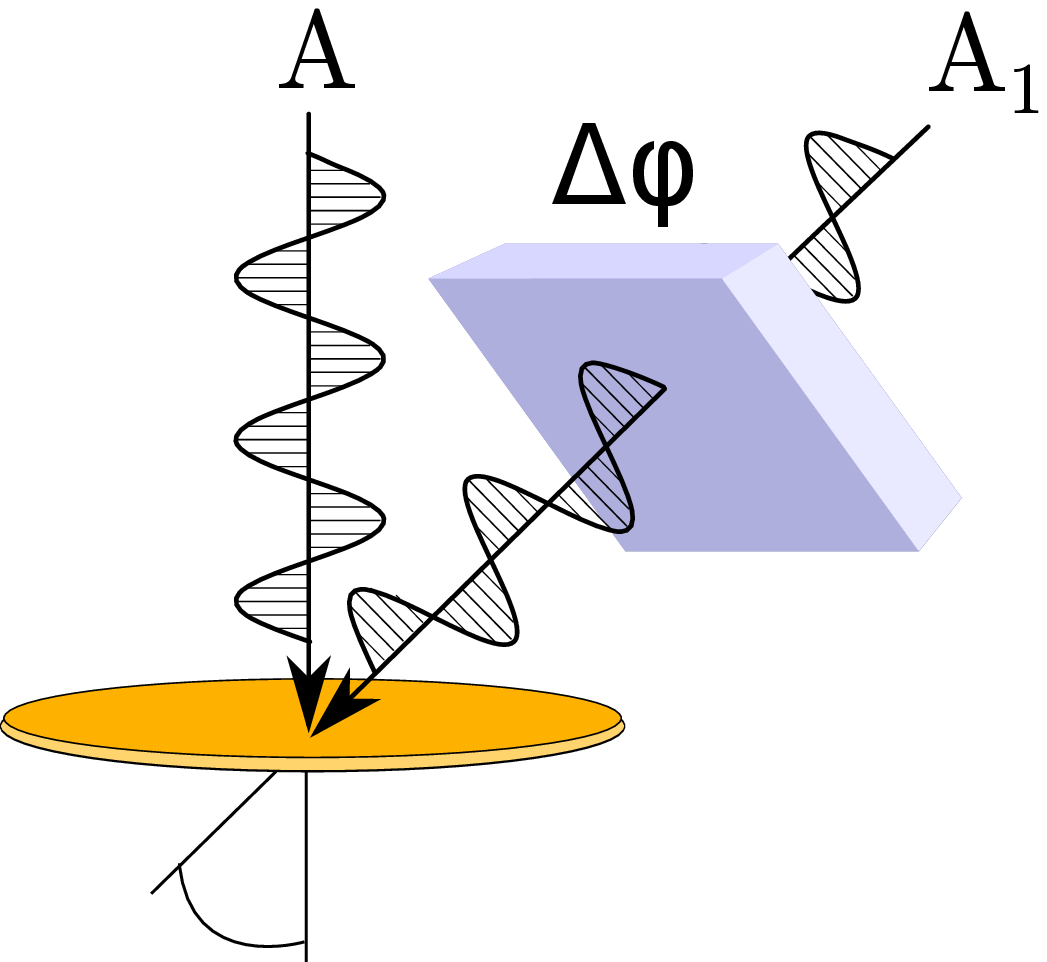}
    b)
    \includegraphics[clip=true,width=.25\textwidth]{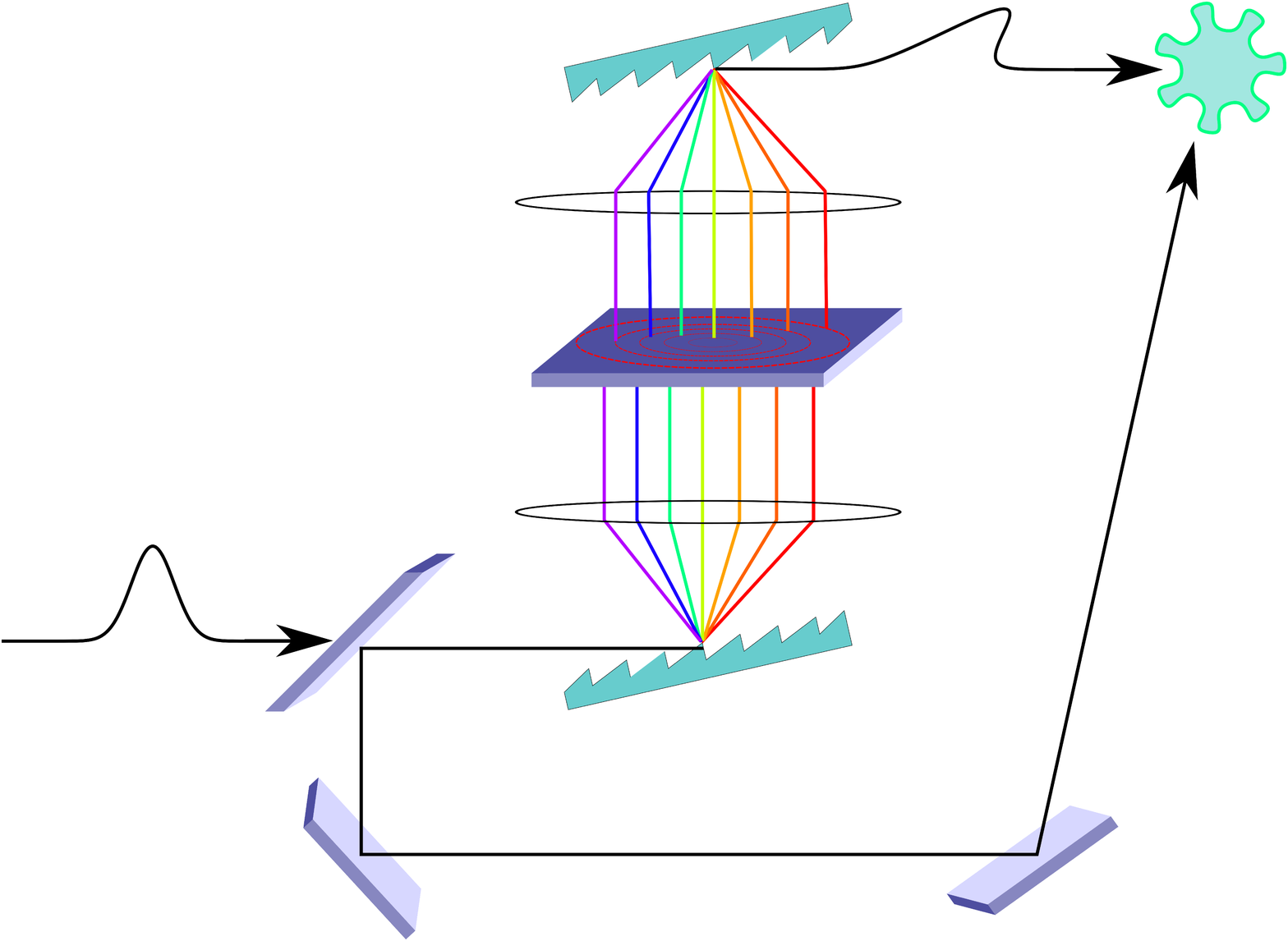}
    c)
    \includegraphics[clip=true,width=.25\textwidth]{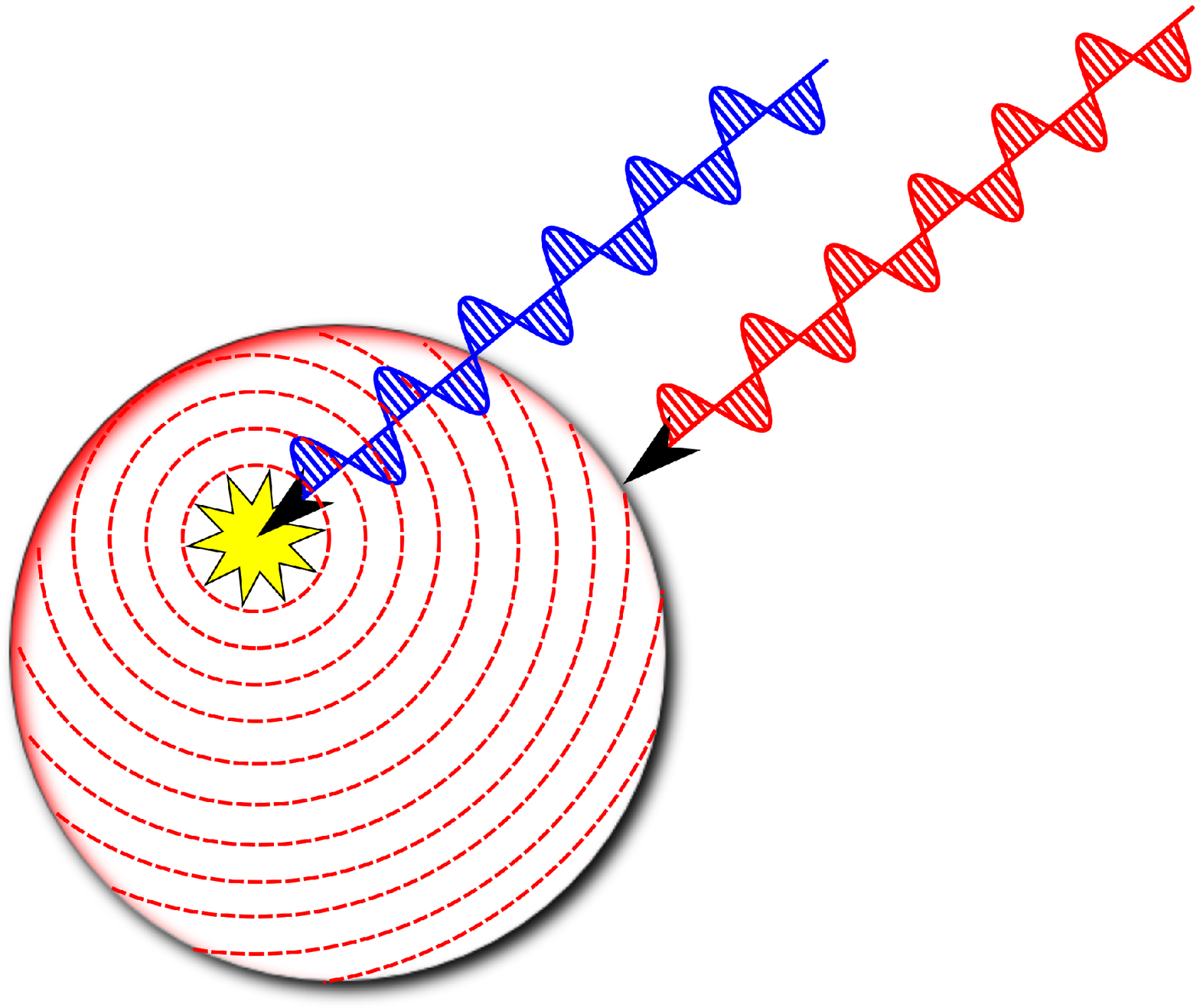}
  \end{center}
  \caption{\label{fig:setups} Suggested experimental geometries for
    control of optical channels. a) Geometry of monochromatic incident
    light fields approaching a disc shaped nanostructure to change the
    amplitudes of a specified principal mode, independent light
    sources with amplitudes $A$ and $A_1$ and a specified relative
    phase $\Delta \Phi$ are shown. b) A more general spatial-light
    modulator approach for changing phase and/or amplitude over a
    range of wavelengths for a broad-band source or applied pulse of
    light. c) Generation of phase controlled light from an internal
    source (driven in the blue) through a coherent process such as two
    photon decay or parametric down conversion, with external control
    light incident at the emission wavelength (in red). By exchanging
    red and blue, the same scheme can be used for second harmonic
    generation.}
\end{figure}

In this paper we develop a general analytical theory for the control
of the modes of scattered and internal fields in nanostructures of any
shape and at any frequency which allow us to either enhance or
suppress internal and/or scattered fields and determine their spatial
distribution or angular momentum. Most importantly, we can design the
lineshape of the particle and introduce very narrow features in its
spectral response. This method requires varying the relative
amplitudes and phases of $N+1$ incident fields in order to control $N$
channels.  Modes of the internal and scattered fields of nanoparticles
are coupled pairwise, each pair forming an interaction channel for the
incident light~\cite{papoff11a}.  Some of the scattering modes
efficiently transport energy into the far field, while others are
mostly limited to the near field region around their
nanostructure~\cite{doherty13a}.  Depending on the channels involved,
it is possible to determine the flow of energy outside the particle by
controlling the scattered field in the far and/or in the near field
regions, or the absorption of energy by the particle through
controlling the amplitude of the internal field.  The incident fields
considered here are commonly available in experiments and can be
monochromatic or broad band.  A practical implementation requires
simply the control of the relative phases of incident fields and can
be achieved using only one source of light together with beam
splitters and phase modulators, or a mixture of sources, which may
also be impurities embedded in the nanoparticle, such as atoms,
molecules or quantum dots. We provide simple sketches of experimental
set-ups suitable to implement our theory in Figure~\ref{fig:setups}.
The coherence length of such sources has to be of the order of the
size of the particle, so even conventional lamps may be used in some
applications, as long as the difference between the optical paths from
the sources to the particle is within the coherence length.  The
internal sources may emit radiation at the frequency under control not
only through elastic scattering, but also through inelastic scattering
and nonlinear processes such as harmonic generation and amplification
of light at the nanoscale. Therefore our approach can be used to also
optically control non-linear processes. In the following we review the
theory of internal and scattering modes for smooth particles, derive
the control conditions for fields at the surface of a particle,
consider the types of light sources able to implement those conditions
and show some numerical illustrations of these ideas. We also show
that a simple parameter scan is sufficient to find the optimal control
condition even without a detailed knowledge of the modes of the
particles, as it is the case in most practical applications.

\begin{figure}
  \begin{center}
   \includegraphics[clip=false,width=.45\textwidth]{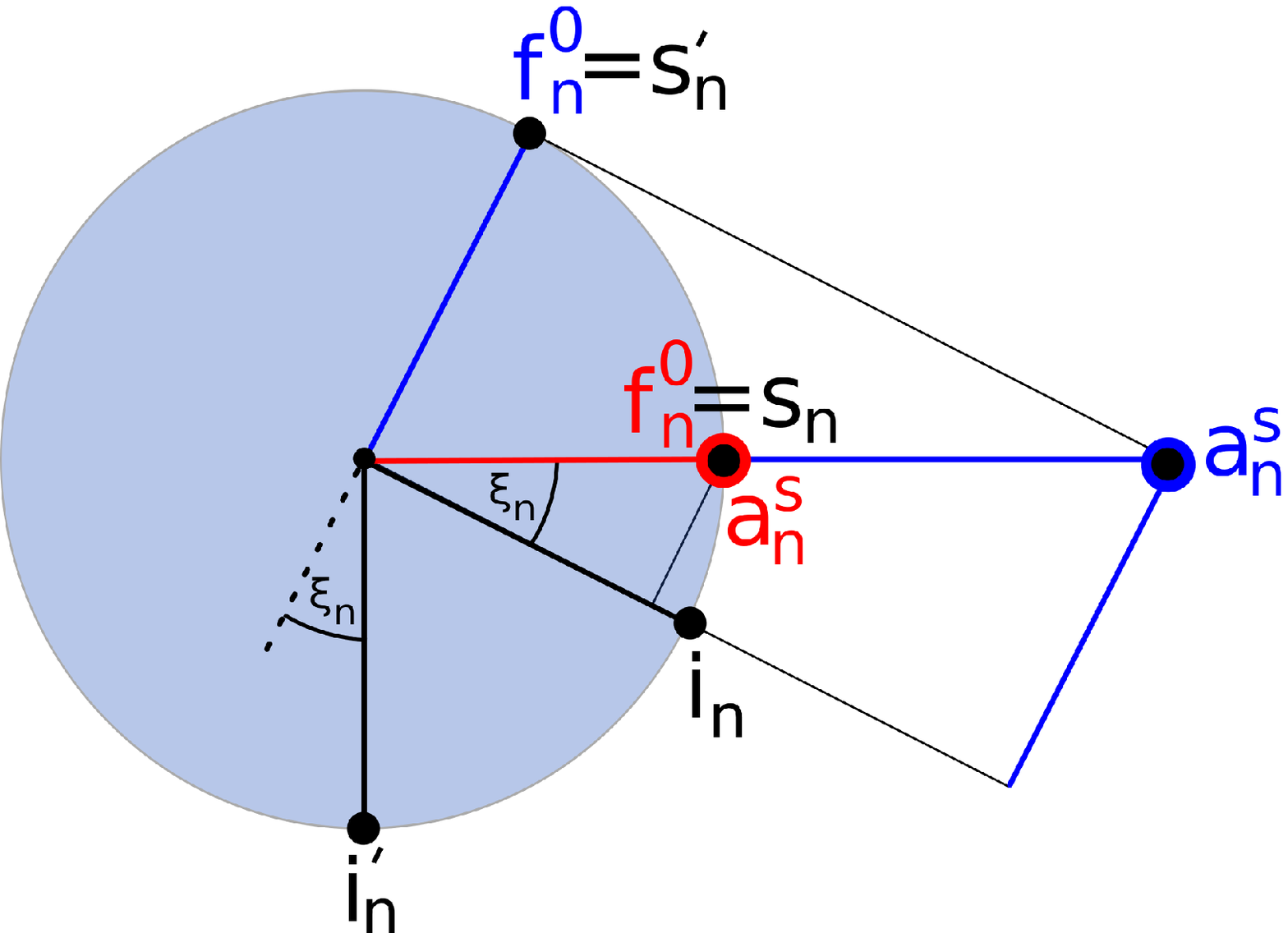}
  \end{center}
  \caption{\label{fig:vectors} Induced amplitudes, $a_n^s$, of the
    $n^\mathrm{th}$ principal modes of a nanostructure, due to fields
    with projections parallel to $s_n$ (shown in red, corresponding to
    Eqn.~\ref{eqn:single_mode_exc}) and $s^\prime_n$ (blue, matching
    Eqn.~\ref{eqn:max_exc}), all modes and incident fields have unit
    magnitide (marked circle). }
\end{figure}

We can generalize Mie theory to non-spherical metallic and dielectric
particles, provided that the particle does not posses sharp edges, and
expand the electromagnetic field at any point in space (both inside
and outside the structure) in terms of the intrinsic modes of the
particle. These modes are in general combinations of different
electric and magnetic multipoles~\cite{papoff11a}. As with the Mie
modes of a sphere, the most important property here is that the
projection on the surface of the particle of each scattering mode is
spatially correlated to the projection of only {\em one} internal
mode, and vice versa, where the spatial correlation is defined by an
integral over the particle surface for the scalar product of the two
projections (see Supplemental Material). Because of this property, the
induced amplitudes of the $n^\mathrm{th}$ pair of principal modes of a
nanostructure can be determined without need of noisy numerical
inversion, by using the coupling of the surface projections of an
incident field ($f$) to the internal ($i_n$) and scattered ($s_n$)
mode pair as
\begin{equation}
  a^i_n = \frac{i^\prime_n \cdot f }{i^\prime_n \cdot i_n} , \; a^s_n
  = - \frac{s^\prime_n \cdot f }{s^\prime_n \cdot s_n},
  \label{eqn:gen_mie}
\end{equation}
here, $g \cdot h$ indicates the surface integral of the scalar product
of the projected fields, $g$ and $h$.  The sub-spaces of the internal
and scattered modes are both orthonormal, i.e., $i_n \cdot i_m = s_n
\cdot s_m= \delta_{nm}$ (with $\delta$ being the Kronecker delta),
there are also associated {\em dual} modes:
\begin{equation}
  i^\prime_n = \frac{i_n - (i_n \cdot s_n) s_n}{\sqrt{1 - (i_n \cdot
      s_n)^2}}, \; s^\prime_n = \frac{s_n - (i_n \cdot s_n)
    i_n}{\sqrt{1 - (i_n \cdot s_n)^2}}
\end{equation}
where $i^\prime_n \cdot i^\prime_m = s^\prime_n\cdot s^\prime_m=
\delta_{nm}$, $i^\prime_n \cdot s_n = s^\prime_n \cdot i_n = 0$ and
$i^\prime_n \cdot i_n = s^\prime_n \cdot s_n$. Note that in this work
that $i^\prime_n$ and $s^\prime_n$ are chosen to be unit vectors, as
distinct from Ref.~\cite{papoff11a}.  The spatial correlations between
modes that appear in this theory can be interpreted geometrically, by
introducing the principal cosines and sines, $i_n \cdot s_n
=\cos{(\xi_n)} $ and $i^\prime_n \cdot i_n = s^\prime_n \cdot s_n
=\sin{(\xi_n)}$. The relative orientation of the $n^\mathrm{th}$ set
of modes are shown schematically in
Figure~\ref{fig:vectors}. Correlations, princple vectors and
coefficients $\{a^{i,s}\}$ depend on the wavelength parametrically
through the frequency dependance of the dielectric and magnetic
permittivity and permeability~\cite{papoff11a}.

From Eq.~\eqref{eqn:gen_mie}, one can see~\cite{papoff11a} that strongly
correlated, i.e., closely aligned modes are at the origin of
resonances in nanoparticles as well as large transient gain and excess
noise~\cite{firth05a} in unstable optical cavities and classical or
quantum systems governed by non-hermitian
operators~\cite{papoff08a,papoff12a}.

General geometrical considerations~\cite{farrell96a} allow us to
determine the surface fields which excite $s_n$ but not $i_n$ (or
$i_n$ and not $s_n$), or that produce the largest amplitude for modes
$s_n$ or $i_n$. We recall that any incident field can be decomposed as
$f=f_n + f_{n_\perp}$, with $f_n $ the part of the incident field that
couples only with the $n^\mathrm{th}$ modes and $f_{n_\perp}$ being
the part that does not.  We find that
\begin{eqnarray}
  & & f_n = s_n \rightarrow a^s_n= 1, \; a^i_n =
  0 \label{eqn:single_mode_exc} , \\ & & f_n = s^\prime_n \rightarrow
  a^s_n= \frac{1}{\sin{(\xi_n})}, \; a^i_n= -
  \frac{\cos{(\xi_n)}}{\sin{(\xi_n})}.\label{eqn:max_exc}
\end{eqnarray}
Eqn.~\eqref{eqn:single_mode_exc} give the requirement for incident
fields that, irrespectively of $f_{n_\perp}$, produce excitation of
only the scattering mode, i.e., null amplitude for the corresponding
internal mode.  Alternatively the largest amplitude of the scattering
mode is obtained for fields with the form of
Eq.~\eqref{eqn:max_exc}. Note that we are considering only incident
fields with $f_n \cdot f_n = 1$ in
Eqs.~(\ref{eqn:single_mode_exc},\ref{eqn:max_exc}) to avoid trivial
effects due to the overall amplitude of the incident fields.
Figure~\ref{fig:vectors} depicts the corresponding incident fields and
the associated amplitudes of the scattered light in mode $s_n$ for
both types of incident field. The analogous conditions for $i_n$ are
found by exchanging $s$ with $i$ in Eqs.~(\ref{eqn:single_mode_exc},
\ref{eqn:max_exc}).

\begin{figure}
\begin{center} a)
   \includegraphics[clip=true,width=.9\textwidth]{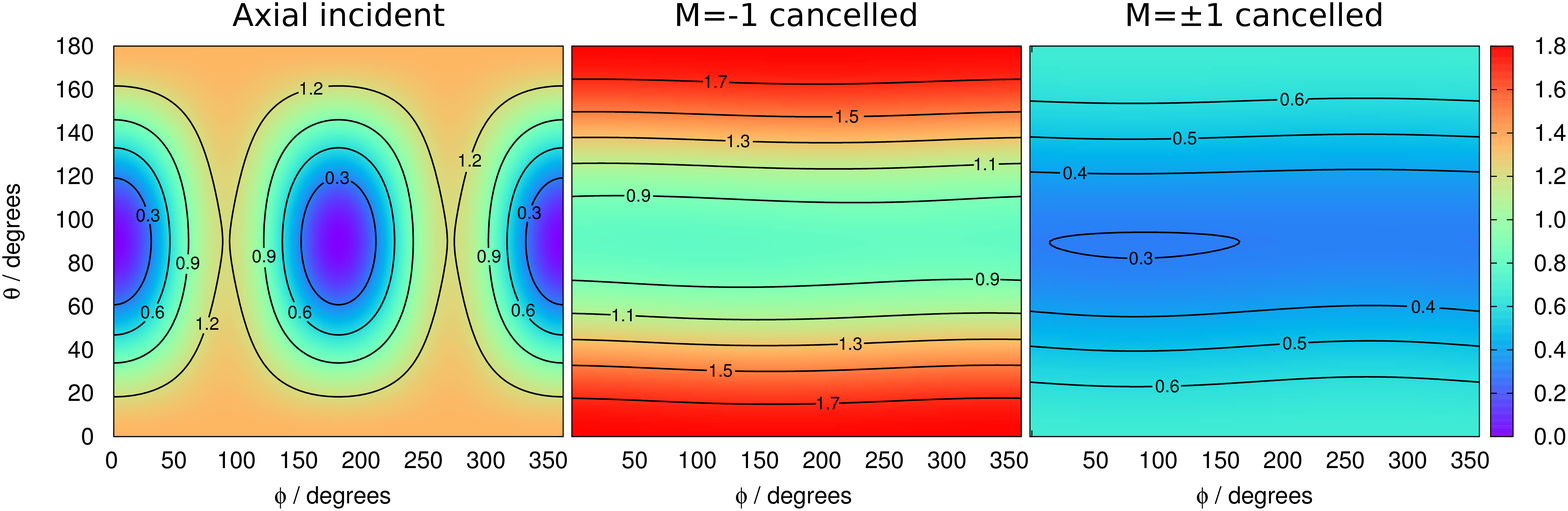}
\end{center}   
 
\begin{center} b)
   \includegraphics[clip=true,width=.7\textwidth]{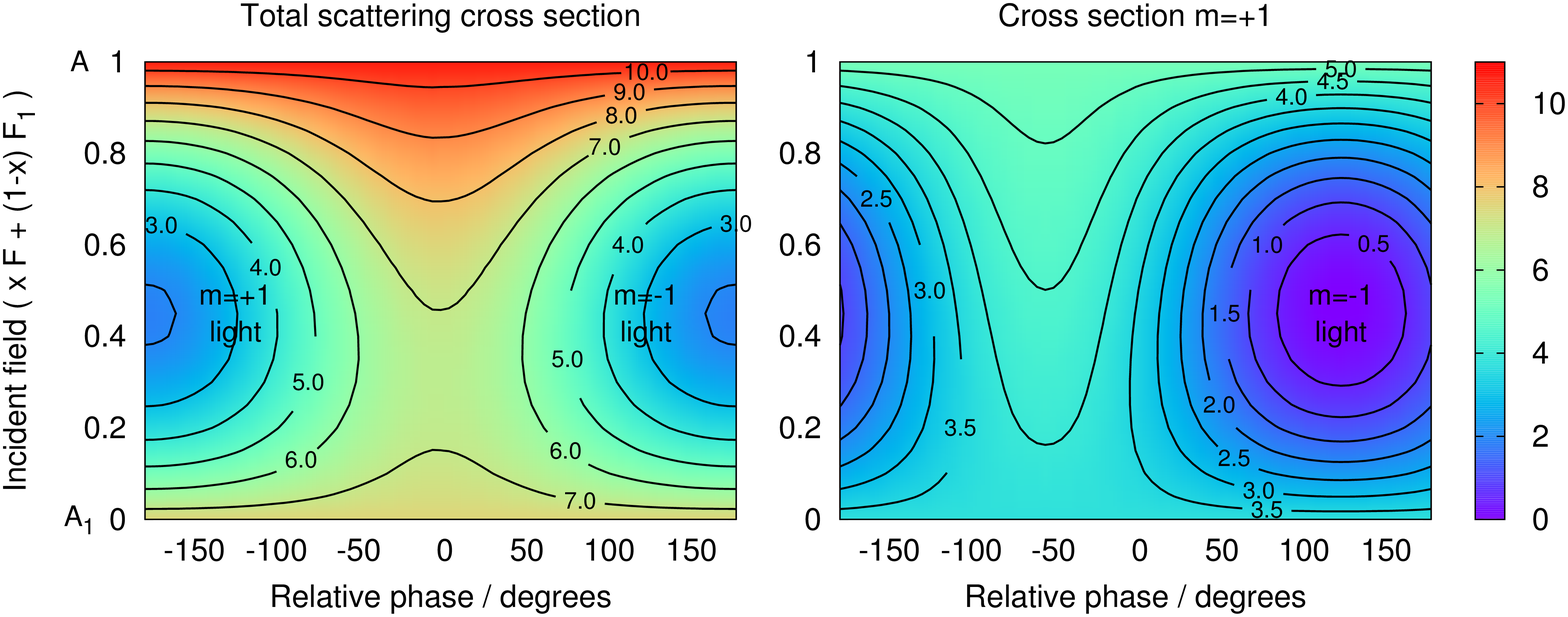}
\end{center}
  \caption{\label{fig:DSCS} Differential scattering for a rounded disc
    shaped (20~nm thickness and 60~nm radius) gold disc, with a
    dominant single pair mode modes showing a dipolar resonance at
    697~nm (in a surrounding medium~\cite{Malitson65} with
    $\epsilon_r$ scaled by 0.75 to emulate support on a silica surface
    in air.) a) Scattering from an incident plane wave clearly shows a
    dipole radiation pattern, but this consists light with both
    $m=\pm1$ angular momentum, by introducing a second and then third
    field with chosen complex amplitudes, the light in the $m=-1$ and
    then both $m\pm1$ channels can be removed. b) Illustration of the
    effect of scanning for the condition to remove the $m=-1$ channel
    by adjusting the magnitude and phase of a second incident
    planewave. Experimentally, this example requires monitoring the
    orbital angular momentum of light emitted along the particle axis,
    to determine the relative amplitudes and phase where only the
    $m=-1$ dipolar resonance is active. Here we show the effect of
    scanning the relative amplitude and phase of axial and an
    additional $45^\circ$ direction (with respect to the particle
    axis) plane-polarised light, between being purely the axial
    ($x=1$, field A) and inclined ($x=0$, field $A_1$) incident
    light.}
\end{figure}

Two points are worth noting. First, these are exact conditions for the
surface fields that are valid at any frequency and have two possible
applications: control of a mode (or modes) over a range of frequency
and introduction of narrow band features in the spectral response of
the particle. Second, the largest amplitude for a mode is not achieved
through single mode excitation, but by an optimal excitation that
produces amplitudes in the two principal modes of the channel. For
physical applications it is necessary to generate incident fields
which have tangent components that fulfill
Eqs.\eqref{eqn:single_mode_exc}, or~\eqref{eqn:max_exc}. The case
$f=s_n$ can be in principle realized through time reversal of the
lasing mode of an amplifier with the same shape as the particle and
gain opposite to the loss~\cite{noh12a} at the resonant frequency of
the mode.  To experimentally realize this is quite challenging, as
radiation would need to converge towards the particle from all
directions.  We are instead interested in deriving general conditions
for optimal excitation of modes at any frequency with easily
accessible sources of radiation. In general, an incident field
$F(\mathbf{r})$ with tangent components $f=i_n$ or $f=s_n$ cannot be
realized using common sources of radiation external to the particle,
such as laser beams or SNOM tips, or even internal sources such as
fluorescent or active hosts. This is because these sources emit waves
that are neither outgoing radiating waves in the external medium, as
is the scattering mode, $s_n$, nor standing waves in the internal
medium, such as the internal mode, $i_n$.  However, by combining two
or more of these sources with appropriate phases and amplitudes, it is
possible to control in a simple and effective way the few dominant
interaction channels of any nanoparticle. To construct fields to
realise the conditions of Eqs.~\eqref{eqn:single_mode_exc}
or~\eqref{eqn:max_exc} requires two linearly independent incident
fields, $A f$, and $A_1 f^1$, both coupled to the channel $n$, such
that
\begin{equation}
  \frac{s_n \cdot f}{s_n \cdot f^1} \ne \frac{i_n \cdot f}{i_n
    \cdot f^1}. \label{eqn:frac}
\end{equation}
The conditions for the suppression of $i_n$ or maximal excitation of
$s_n$ are realized, up to a scale factor, by choosing $ A_1$ as
follows:
\begin{eqnarray}
  A_1 = - A \frac{i_n \cdot f}{i_n \cdot f^1} ,
  \label{eqn:max_exc_2}\\
  A_1  = - A \frac{i^\prime_n \cdot f}{i^\prime_n \cdot
    f^1} , \label{eqn:mode_supp}
\end{eqnarray}
with $A$, $A_1$ being complex amplitudes that can be experimentally
adjusted through phase plates and dicroic elements
(Figure~\ref{fig:setups} suggests a geometry for constructing these
fields, while Figure~\ref{fig:DSCS}b) shows a numerical example of
scanning the relative complex amplitude to produce specific scattered
light).  Analogous conditions for optimization of $i_n$ and
suppression of $s_n$ can be found by swapping $i_n, i^\prime_n$ with
$s_n, s^\prime_n$ in Eqs.~(\ref{eqn:max_exc_2}, \ref{eqn:mode_supp}):
the generalization of Eq.~\eqref{eqn:max_exc_2} to a larger number of
modes is presented in the Supplemental Material.

We note that conservation of energy applies to the incident scattered
and internal fields, but not necessarily to each interaction channel
separately.  However, if incident fields with $f_n \ne 0, f_{n_\perp}
= 0$ exist, the conservation of energy applies to the $n$ channel.
From the Stratton-Chu representations~\cite{stratton39a} (see
Supplemental Information) we can show that in particles where the
dependence of the electric (and magnetic) parts of $i_n,s_n,f_n$ on
the surface coordinates is the same, such as for sphere and cylinders,
the ratios in Eq.~\eqref{eqn:frac} depends on the flux of energy of
the incident field into the particle, and Eq.~\eqref{eqn:frac} is {\em
  always} violated when the sources are both either outside or inside
the particle. For instance, explicitly in the case of a sphere any
external source can be expanded by the set of regular multipoles with
angular indexes $l,m$, while any internal source is expanded by the
set of radiating spherical multipoles, as shown in the Supplementary
Material. In this case the amplitudes of internal and scattering modes
cannot be controlled by independent external -- or internal --
sources: Eqs.~(\ref{eqn:max_exc_2},\ref{eqn:mode_supp}) become
equivalent and lead to the simultaneous suppression of both $i_n$ and
$s_n$, while the maximization of the amplitudes of both scattering and
internal modes is provided by ensuring that the contributions of the
incident fields to the amplitudes add in phase.  However, maximal
excitation, Eq.~\eqref{eqn:max_exc_2}, or suppression of a mode can be
obtained also in these particles at any frequency provided one source
is external and the other internal, or that one incident field is a
regular wave with with a power flow of $W_n = 0$ and the other an
incoming wave with $W_n \ne 0$.  Examples of internal sources
important for applications are an impurity scattering inelastically at
the same frequency as the external control beam or an active center
excited non radiatively.  Incoming waves are more difficult to
realize, but could be in principle be obtained through time reversal
techniques~\cite{pierrat13a}.

\begin{figure}
\begin{center}
   \includegraphics[clip=true,height=.40\textheight]{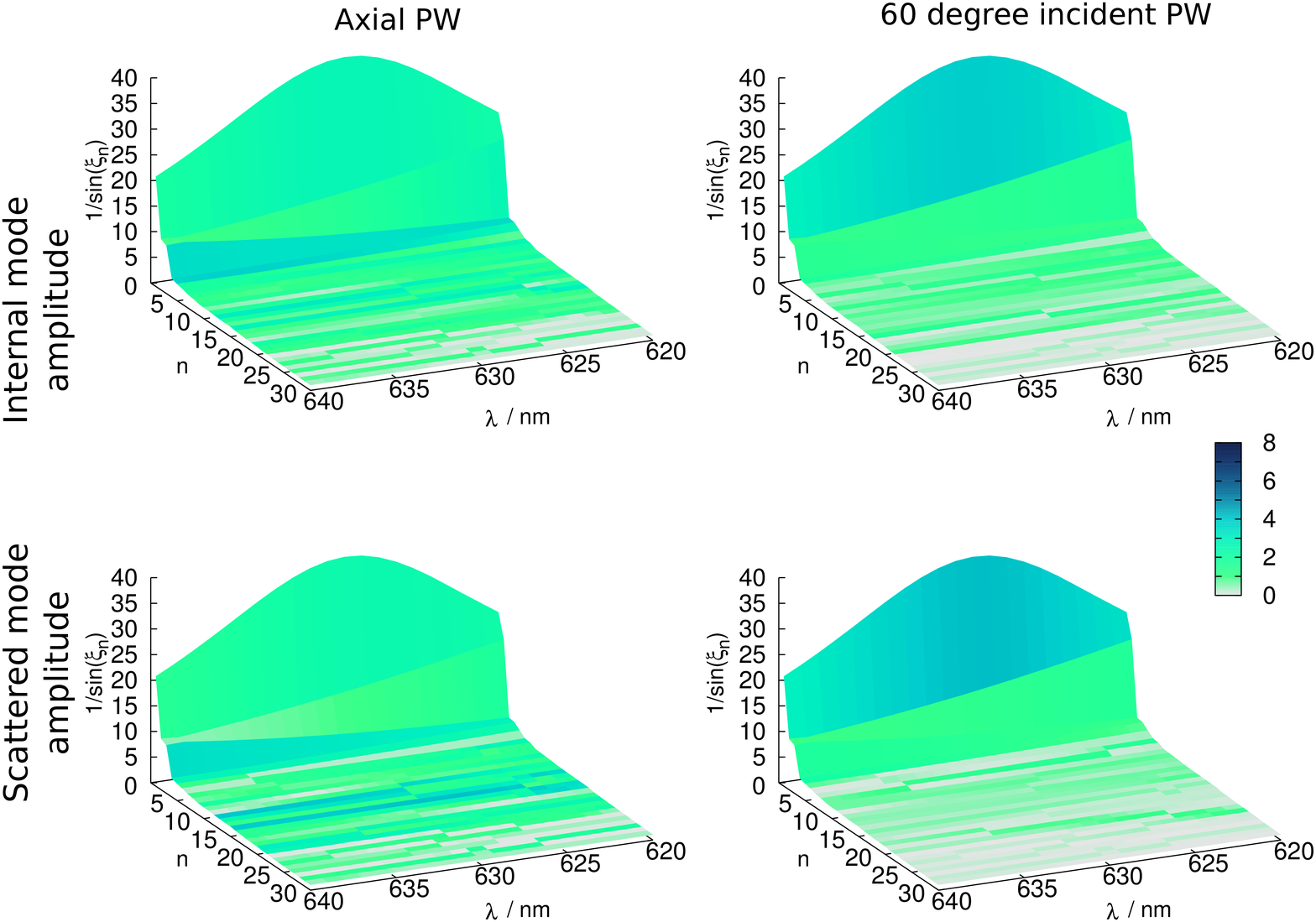}
   \includegraphics[clip=true,height=.35\textheight]{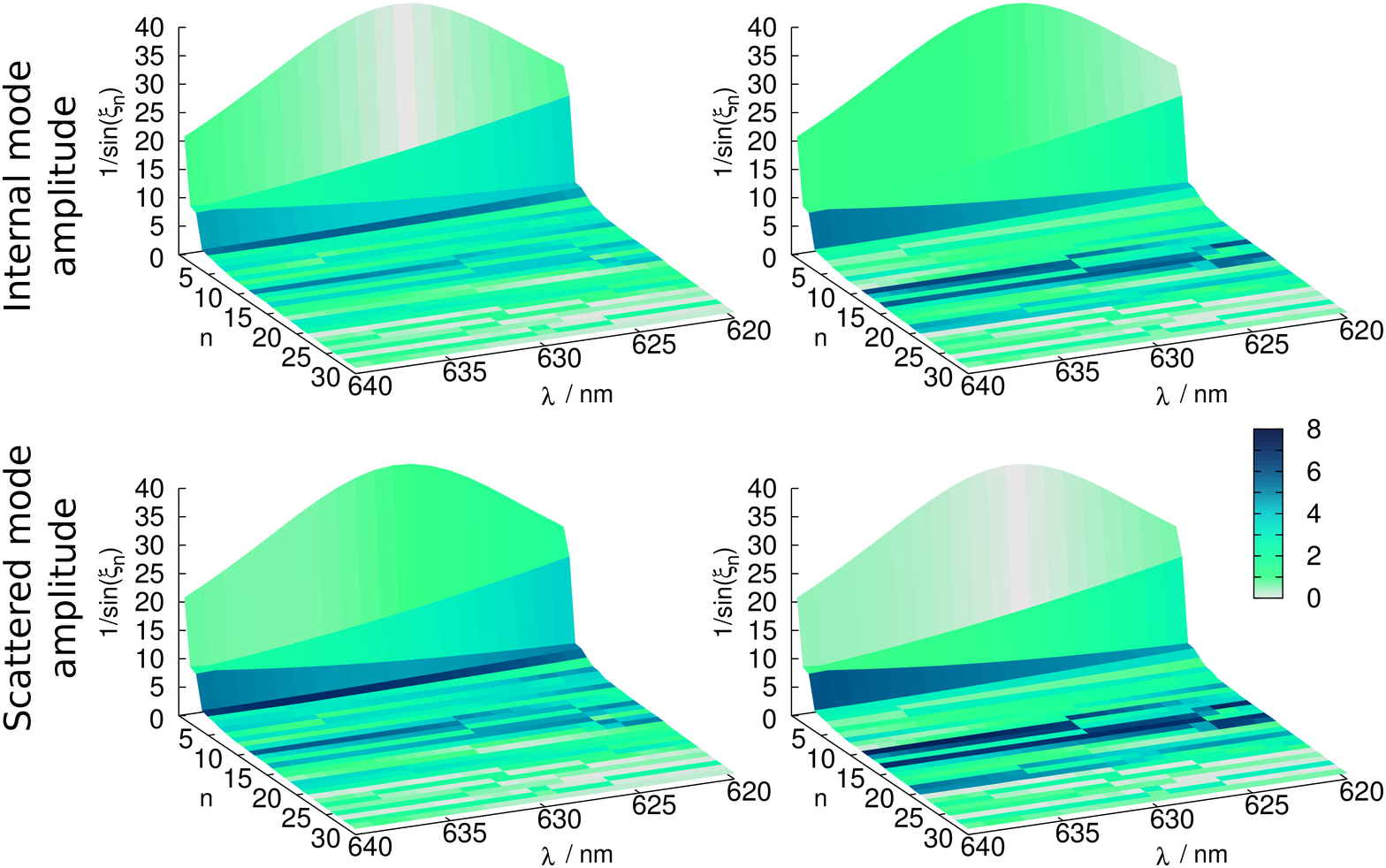}
\end{center}   
\caption{\label{fig:landscapes} Internal and scattered mode amplitudes
  of a rounded rod-shaped dielectric ($n=1.5$) particle, of 2~$\mu$m
  length and 0.7~$\mu$m diameter around one of its (many) resonances
  at $\sim 630$~$nm$. The height of this landscape is
  $\sin^{-1}({\xi})$, i.e.\ the largest value of amplitude possible
  for $|f_n|=1$; while the shading of the traces overlaid on top show,
  for each wavelength ($\lambda$), the amplitude induced by the
  incident combinations for the first $n$ most aligned principle
  modes. For fields which obey Eqn.~\eqref{eqn:frac}, the internal and
  scattered modes can be independently addressed. Here we use a second
  field with $60^\circ$ direction of incidence and polarisation,
  chosen to either cancel the peak of the resonant feature at the back
  of the landscape for the internal or scatted cases respectivly
  (leading to the disapearance of amplitude in only that particular
  mode).}
\end{figure}

\begin{figure}
  \begin{center}
    \includegraphics[clip=true,width=.75\textwidth]{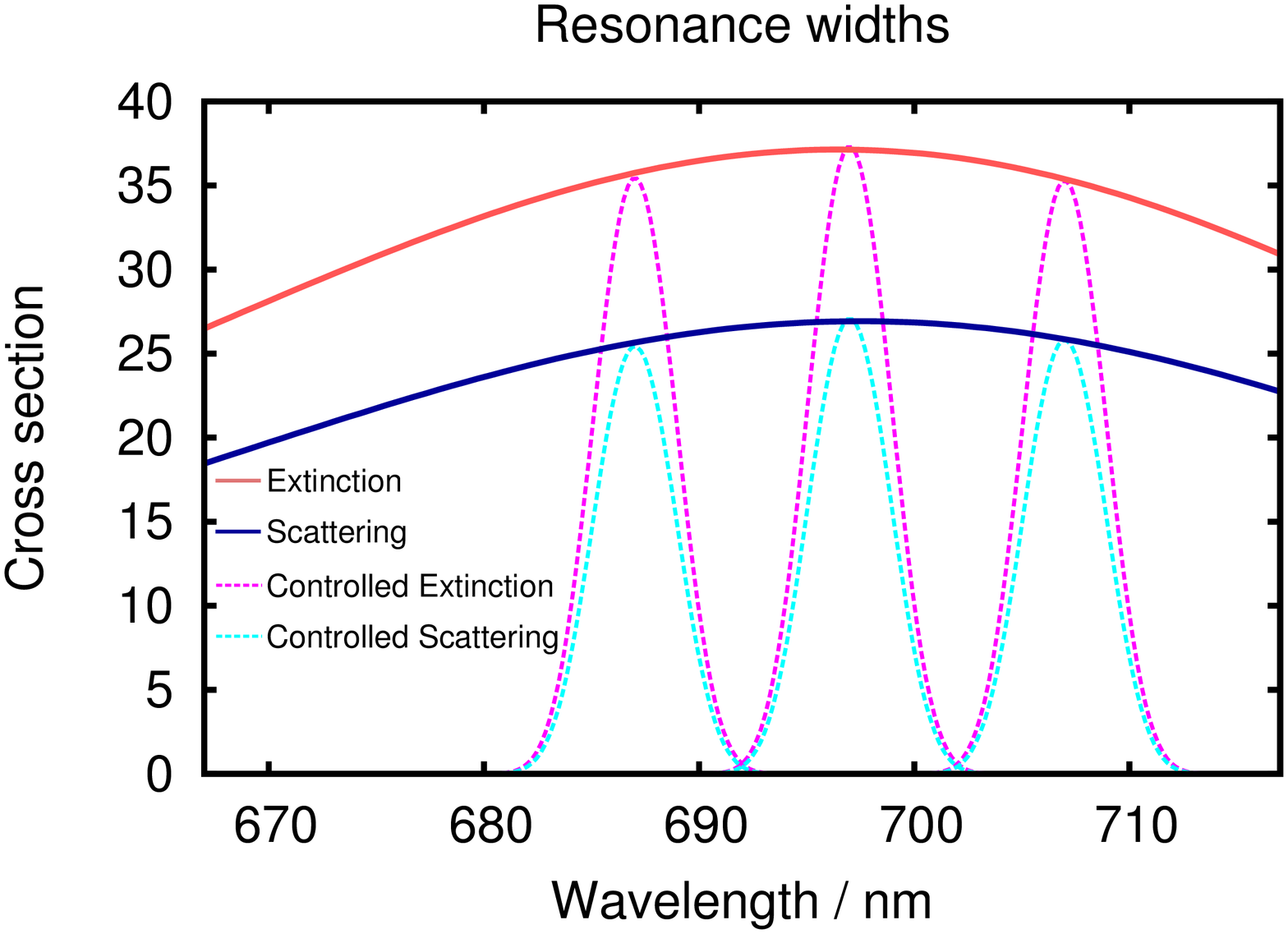}
  \end{center}
  \caption{\label{fig:sharplines} Cross sections of the gold disc from
    Figure~\ref{fig:DSCS} with standard axial incidence, leading to a
    broad feature in the extinction (red) and scattering (blue), this
    is compared again its response to illumination of light from three
    directions (axial, $45$ and $90^\circ$ incident) with their
    relative phases (but not amplitudes) modulated to cause
    constructive interference only at a chosen wavelength, then
    rotating the phase to cause destructive interference according to
    a gaussian envelope, producing a specified location and line width
    feature within the envelope of the original broad peak.}
\end{figure}

We now show numerically how the theory developed here can be applied.
In Figure \ref{fig:landscapes} we show how we can suppress either the
scattered or the internal mode of the resonant channel in a rounded
rod-shaped particle, introducing in this way features that are much
narrower than the resonance. The height and the position of this
curves, the "mode landscape"~\cite{papoff11a} are a property of the
particle, but the amplitude of the modes, which is color coded in the
figure, depends also on the incident fields. The width of the narrow
structure depends on the spectral resolution of the elements used in
the set-up shown in Figure \ref{fig:setups} b.  This application leads
to a much higher spectral resolution in surface enhanced spectroscopy
and sensing.  In Figure \ref{fig:DSCS} we show how the angular
momentum of the scattered light can be controlled in a gold disc. The
disc has an electric dipole radiation pattern and by suppressing the
channels with $m=\pm 1$, where $m$ is the eigenvalue of the angular
momentum around the axis of the disc, we can make the disc virtually
invisible in the far field. Using the same technique over a range of
wavelengths we can control and manage the dipole resonance of this
disc, introducing lines that have a full width half maximum one order
of magnitude smaller than the width of the dipole resonance, see
Figure \ref{fig:sharplines}.

In conclusion we have shown that spatial distribution and angular
momentum of internal and scattered fields of dielectric or metallic
nanoparticles can be configured and optimized by controlling the
amplitudes of a finite number of internal and scattered modes of the
nanoparticle through the relative phases of coherent sources of
light. This can be achieved using a single source of light of the type
normally available in experiments, beam splitters and phase
modulators, or combination of internal and external sources. The
number of sources required is equal to the number of channels to
control plus one, so this general approach is particularly simple to
implement near resonances, where the response to light of
nanoparticles is dominated by few modes.  We have also been able to
use optical control to reshape the spectral response of nanoparticles,
introducing narrow spectral lines of a few nanometers width. In
general, the width of the narrow lines depends on the spectral
resolution of the control set-up, i.e. of the gratings and of the
diffractive elements used and can be much narrower than the resonance
peaks of the nanoparticle. This can significantly improve the spectral
resolution of surface enhanced spectroscopy, which is based on the
enhancement of surface fields near nanoparticles. The same control
approach can be applied to light generated through linear and
nonlinear processes by impurities embedded in the particle. In this
case, controlling internal modes allows one the management of
non-linear transitions in the impurities.  Most of all, the potential
of this work lies in its generality and in the fact that it does not
requires complex spatial shaping of the incident fields, but only the
control of their relative phases and amplitudes that can be provided
by already available diffractive elements such as spatial light
modulators.





\clearpage
%
%
\section*{Supplemental Information}
We consider metallic and dielectric particles without sharp edges and
use~\cite{holms09a} vectors, $F = [E,H]^T$, with three electric and
three magnetic components for the electromagnetic fields; the
corresponding surface field, $f$, is made up by the two electric and
two magnetic components of $F$ tangent to the surface of the particle.
In this notation, the boundary conditions are $f=f^i-f^s$, i.e. the
tangent components $f$ of the incident field are equal to the
difference between the tangent components of internal and scattered
fields, $f^i$ and $f^s$ at the surface.  We can find solutions of the
Maxwell 's equation in the internal and external media, the principal
modes ${I_n},{S_n}$, such that the expansion of the scattered and
internal fields converge to the exact fields for any incident
field~\cite{rother02a,holms09a, papoff11a}. The tangent components of
the principal modes, $\{i_n\}$ and $\{s_n\}$, form bases for the
internal and scattered surface fields that are orthonormal with
respect to the scalar product $f \cdot g = \sum_{j=1}^4 \int_S f_{j}^*
g_{j} ds$, where $j$ labels the components of the surface fields and
the asterisk stands for the complex conjugate.  Each pair $i_n,s_n$
define an interaction channel for the particle that is independent
from all the other pairs because each internal mode $i_n$ is
orthogonal to all but the scattering mode $s_n$ and viceversa.  The
spatial correlations that appear in this theory can be interpreted
geometrically. The principal cosine between $s_n$ and $i_n$ is defined
by the overlap integral, or correlation~\cite{hannan61a}, as
$\cos{(\xi_n)} \equiv i_n \cdot s_n$, with the arbitrary phases of
$i_n,s_n$ chosen so that the integral is either positive or null.  The
corresponding function $\sin{(\xi_n)}$ is the orthogonal distance
between $s_n$ and $i_n$.  The angles ${\xi_n}$ characterize the
geometry of the subspaces of the internal and scattered solutions and
are invariant under unitary transformations~\cite{knyazev10a}.

Given the importance of spheres in scattering, we show explicitly how
the optimization or suppression of either scattering or internal modes
can be obtained when one source is internal and the other external,
but when not both sources are either internal or external.  The
internal and scattering modes used in this work are proportional to
the usual Mie modes: writing explicitly the electric and magnetic
component, the magnetic multipole modes are defined, up to an
arbitrary phase factor, as
\begin{eqnarray}
i_{lm}= \frac{e^{i\phi_l} }{( | i^E_l(r)|^2 +| i^H_l(r)|^2)^{1/2}}
[m_{lm}(\Omega) i^E_l(r) , n_{lm}(\Omega) i^H_l(r)]^T,
\nonumber\\ s_{lm}= \frac{1 }{ (|s^E_l(r)|^2 +|s^H_l(r)|^2)^{1/2}}
	    [m_{lm}(\Omega) s^E_l(r), n_{lm}(\Omega) s^H_l(r)
	    ]^T \label{Eqn:Mie_modes}
\end{eqnarray}
where $r$ is the radius of the sphere, $\Omega$ the solid angle, $k_i,
k$ are the wavenumbers for internal and external medium, respectively,
$m_{lm},n_{lm}$ vector spherical harmonics. $ i^E_l(r)=j_l(k_ir)$, $
i^H_l(r)=-i(\epsilon_i/\mu_i)^{1/2} (k_ir)^{-1} \partial_{k_ir} k_ir
j_l (k_ir)$, $s^E_l(r)=h_l(k r)$, $s^H_l(r)=-i (\epsilon/\mu)^{1/2} (
kr)^{-1} \partial_{k r} k r h_l (k r)$, $j_l, h_l$ the spherical
Bessel and Hankel functions of order $l$.  $\phi_l= -Im\{ \log{[
    {i^E_l(r)}^*s^E_l(r) + {i^H_l(r)}^*s^H_l(r)]}\}$ is such that
\begin{equation}
i_{lm} \cdot s_{lm} = \frac{{|i^E_l(r)}^*s^E_l(r) +
  {i^H_l(r)}^*s^H_l(r)|}{( | i^E_l(r)|^2 +|
  i^H_l(r)|^2)^{1/2}(|s^E_l(r)|^2 +|s^H_l(r)|^2)^{1/2}}.
\end{equation}
Electric multipole modes are obtained as usual exchanging magnetic
with electric parts. The incident fields generated by two external
sources can be expanded in terms of electric and magnetic multipoles
centered on the sphere; the only terms of these expansions that can
couple to the modes in Eqs.~\eqref{Eqn:Mie_modes} are $f= a
[m_{lm}(\Omega) f_l^E(k r), n_{lm}(\Omega) f_l^H(k r)]^T$, $f^1= a_1
[m_{lm}(\Omega) f_l^E(k r), n_{lm}(\Omega) f_l^H(k r)]^T$, where
$f_l^E(k r),f_l^H(k r)$ are found by replacing $k_ir$ with $kr$ in
$i^E_l,i^H_l$ . Therefore we find that both terms in
Eq.~\eqref{eqn:frac} are equal to $a/a_1$, i.e. the inequality is
violated. On the contrary, the equality can be satisfied when the two
sources are such that ${f_l^1}^E \ne f_l^E,{f_l^1}^H \ne
{f_l}^H$. This is the case when one source is external and the other
is a radiating multipole inside the particle with $f^1= a_1
[m_{lm}(\Omega) {f_l^1}^E(k_i r), n_{lm}(\Omega) {f_l^1}^H(k_i r)]^T$,
where ${f_l^1}^E(k_i r),{f_l^1}^H(k_i r)$ are found by replacing $kr$
with $k_ir$ in $s^E_l,s^H_l$. For this combination of internal and
external sources Eqs.~(\ref{eqn:max_exc_2}, \ref{eqn:mode_supp}), for
suppression of $i_{lm}$ and maximal excitation of $s_{lm}$, can be
applied and are
\begin{eqnarray}
A_1 &=& - A \frac{{f_l^E}^* i_l^E + {f_l^H}^* i_l^H}{{f_l^1}^{E*}
  i_l^E + {f_l^1}^{H*} i_l^H}, \\ A_1 &=& - A \frac{{f_l^E}^*(
  e^{i\phi_l}i_l^E- \cos{(\xi_n)} s_l^E) + {f_l^H}^*
  (e^{i\phi_l}i_l^H- \cos{(\xi_n)} s_l^E)}{ {f_l^1}^{E*} (e^{i\phi_l}
  i_l^E- \cos{(\xi_n)} s_l^E) +{f_l^1}^{H*}(e^{i\phi_l}i_l^H-
  \cos{(\xi_n)} s_l^E)}.
\end{eqnarray}
 The same analysis holds true also for infinite cylinders.
 
We consider here when an field generated by external source with $f_n,
\ne 0, f_{n_\perp} = 0$ exist.  Because $i_n,s_n,f_n$ are exact
solutions of the Maxwell's equations, the Stratton-Chu
representation~\cite{stratton39a} for particles with continuously
varying tangents (particles of class $C^1$), and the corresponding
jump relations~\cite{doicu00} at the boundary of the particle
 
 \begin{eqnarray}
 \int \nabla \times (n(y) \times i_n(y)) g(x,y,k_i) da(y) =
 \frac{1}{2} n(x) \times (n(x) \times i_n(x)) \label{eqn:SCi}\\ \int
 \nabla \times (n(y) \times s_n(y)) g(x,y,k) da(y) = - \frac{1}{2}
 n(x) \times (n(x) \times s_n(x)) \label{eqn:SCs} \\ \int \nabla
 \times (n(y) \times f_n(y)) g(x,y,k) da(y) = \frac{1}{2} n(x) \times
 (n(x) \times f_n(x)) \label{eqn:SCf},
 \end{eqnarray}
 with $x,y$ points on the surface and $n(x)$ the outgoing normal at
 $x$.  Using the Eqs.\eqref{eqn:SCi},\eqref{eqn:SCs} in
 Eq.\eqref{eqn:SCf} and the conservation of energy, we find
\begin{eqnarray}
& & \int \nabla \times (n(y) \times i_n(y)) g(x,y,k) da(y) =
  \frac{1}{2} n(x) \times (n(x) \times i_n(x)) \nonumber \\ & &
  -\frac{a^s_n}{a^i_n}n(x) \times (n(x) \times
  s_n(x)) \label{eqn:SCisf}\\ & & W_n = |a^i_n|^2 W^i_n +|a^s_n|^2
  W^s_n + \nonumber \\ & & \frac{1}{2} Re [|a^i_n||a^s_n| e^{i\Delta}
    \int n(y) \cdot (i_n^E(y) \times s_n^{H*}(y)) da(y) +\nonumber
    \\ & & |a^i_n||a^s_n| e^{i\Delta} \int n(y) \cdot (s_n^E(y) \times
    i_n^{H*}(y))] da(y) = 0, \label{eqn:energy}
\end{eqnarray} 
where $\Delta$ is the phase difference between $a^i_n, a^s_n$ and
$W_n, W^i_n,W^s_n$ the power fluxes of $f_n,i_n,s_n$.  Because
$|a^i_n|,|a^s_n|$ and $\Delta$ do not depend on $x$, we can see that
Eqs.\eqref{eqn:SCisf},\eqref{eqn:energy} can be solved only if the
dependence on $x$ in all the terms in \eqref{eqn:SCisf} can be
factored out.  This happens for spheres and cylinders.

The procedure in Eq.~\eqref{eqn:max_exc_2} is generalized to $m$
internal modes and $l$ scattering modes using $m+l$ incident fields,
$A_1f^1, \dots,A_{m+l}f^{m+l}$, each satisying Eq.~\eqref{eqn:frac},
as follows
\begin{eqnarray}
 \left[ \begin{array}{ccc} \frac{i^\prime_1 \cdot f^1 }{\sin{(\xi_1)}}
     & \dots & \frac{i^\prime_1 \cdot f^{m+l} }{\sin{(\xi_1)}}
     \\ \vdots & & \vdots \\ \frac{s^\prime_l \cdot f^1
     }{\sin{(\xi_l)}} & \dots & \frac{s^\prime_l \cdot
       f^{m+n}}{\sin{(\xi_l)}}
 \end{array} \right] 
 \left[ \begin{array}{c} A_1 \\ \vdots \\ A_{m+l}
 \end{array} \right] =
 - A \left[ \begin{array}{c} \frac{i^\prime_1 \cdot f }{\sin{(\xi_1)}}
     \\ \vdots \\ \frac{s^\prime_l \cdot f}{\sin{(\xi_l)}}
   \end{array} \right].
\end{eqnarray}
The solution for the amplitudes is unique if the determinant of the
matrix is not null.

\providecommand*\mcitethebibliography{\thebibliography}
\csname @ifundefined\endcsname{endmcitethebibliography}
  {\let\endmcitethebibliography\endthebibliography}{}

%
%
%
\end{document}